\documentclass[%
 reprint, prl,
showpacs,preprintnumbers,
 amsmath,amssymb,
 aps,
]{revtex4-1}

\usepackage{graphicx}
\usepackage{dcolumn}
\usepackage{bm}

\begin{document}

\preprint{APS/123-QED}

\title{Spatiotemporal organization of energy release events in the quiet solar corona}

\author{Vadim M. Uritsky}
 \altaffiliation[Also at ]{Catholic University of America, Washington DC }
 \affiliation{CUA at NASA Goddard Space Flight Center, Greenbelt MD}%
 \email{vadim.uritsky@nasa.gov}
\author{Joseph M. Davila}%
 \email{joseph.m.davila@nasa.gov}
\affiliation{NASA Goddard Space Flight Center, Greenbelt MD}%

\date{\today}

\begin{abstract}

Using data from STEREO and SOHO spacecraft, we show that temporal organization of energy release events in the quiet solar corona is close to random, in contrast to the clustered behavior of flaring times in solar active regions. The locations of the quiet-Sun events follow the meso- and supergranulation pattern of the underling photosphere. Together with earlier reports of the scale-free event size statistics, our findings suggest that quiet solar regions responsible for bulk coronal heating operate in a driven self-organized critical state, possibly involving long-range Alfv\'{e}nic interactions.


\end{abstract}

\maketitle

The Sun's corona has one of the most violent plasma environments in our solar system. Coronal active regions formed by an intense large-scale magnetic convection in the underlying photosphere produce major explosive events such as X-class flares, coronal mass ejections, and filament eruptions causing dramatic space weather effects. Quiet coronal regions dominated by the magnetic network exhibit smaller-scale but abundant energy release events which are instrumental for bulk coronal heating (see e.g. \cite{karpen95, aschwanden06, solanki06, antiochos07} and refs therein). Understanding the physical mechanism responsible for heating the corona via these bursty events remains one the most important problems in astrophysics \cite{klimchuk06}. 

The directly measured coronal brightenings are unable to supply the energy loss rates of $10^5 - 10^7$ erg cm$^{-2}$ s$^{-1}$ required for the coronal heating \cite{athay66}. This discrepancy implies a collective contribution from a large number of partially unresolved energy release events such as those associated with a localized magnetic reconnection \cite{fuentes10} or resonant wave heating \cite{ofman98}. The appearance of such events in the topologically complex and highly conductive coronal plasma involves cooperative interactions across wide ranges of spatial, temporal and energy scales \cite{uritsky13}. 

Self-organized criticality (SOC) \cite{bak87, bak88} and fluid turbulence \cite{monin75} are two plausible statistical-physical scenarios governing multiscale interactions in the corona. SOC models address the thermodynamic limit of Parker's scenario of nanoflare heating \cite{parker83, parker88} under slow-driving conditions \cite{lu91,charbonneau01}, and reproduce many of the observed coronal statistics \cite{charbonneau01, uritsky13}. High-Reynolds number fluid turbulence in coronal loops  (e.g., \cite{nigro04, buchlin07}) offers an alternative path to these probabilistic signatures. 

The two scenarios can be distinguished based on the statistics of occurrence times of dissipation events \cite{boffetta99,lepreti01}. In paradigmatic SOC models, the times of initiation of energy avalanches are fully random since the probability of an avalanche is given by the ratio between the number of minimally stable states and the total system size \cite{bak87}. As the latter increases, this ratio approaches a constant, giving rise to a Poisson statistics of the event occurrence times. In intermittent turbulence, the occurrence times are organized in multiscale clusters revealing the hierarchy of temporal scales of the underlying fluid dynamics \cite{boffetta99}. While the temporal clustering of energy dissipation is not necessarily related to turbulent flows \cite{sanchez02, paczuski05, wheatland00}, its absence makes a strong case for a SOC-like dissipation mechanism.

Until now, analyses of temporal correlations of solar flares have been focused on high-energy eruptive events resulting in significant increases of the extreme ultraviolet or X-ray emission fluxes \cite{boffetta99}. Such events are typically produced by solar active regions \cite{wheatland00, torok11} and do not reflect the dynamics of bulk coronal plasma residing in the quiet regions. 

In this Letter, we present the first statistical study of the occurrence times of heating events in the quiet Sun, in conjunction with spatial clustering of the events. Using the correlation integral (CI) technique, we show that temporal organization of these events is indistinguishable from stationary Poisson process across the entire observed range of scales ($300 - 5 \times 10^4$ s), and is therefore consistent with paradigmatic SOC models. Clustering of event positions follows the convection pattern of the photospheric supergranulation acting as a spatially distributed driver. Together with earlier reports of power-law distributions of heating events, our results suggest that bulk energy conversion in the solar atmosphere occurs via SOC-like avalanches of magnetic energy dissipation, possibly involving long-range Alfv\'{e}nic interactions. 

We studied coronal images obtained from Solar TErrestrial RElations Observatory Extreme Ultraviolet Imager (STEREO EUVI) \cite{scherrer95} representing the dynamics of a quiet Sun during 17:29:00 05/04/2007 -- 10:58:00 06/04/2001. The overall activity level during the observed interval remained low, with the GOES X-ray fluxes staying below $10^{-8}$ W m$^2$. The 171\r{A} bandpass corresponding to Fe IX and Fe X emission lines was used, with the maximum response at the solar plasma temperature $\sim 9 \times 10^5 $K. We also analyzed a  set of Solar and Heliospheric Observatory Michelson Doppler Imager (SOHO MDI) \cite{scherrer95} magnetograms co-aligned with the STEREO EUVI images. Both image sets were derotated and rebinned down to the spatial resolution $0.6$ arcsec, with the average sampling time 66.1 s. The obtained data cubes contained $770\times500\times952$ points in the latitudinal, longitudinal, and temporal directions, correspondingly. The field of view was close to the disk center ensuring small projection distortions.

\begin{figure}
\includegraphics[width=8.5 cm]{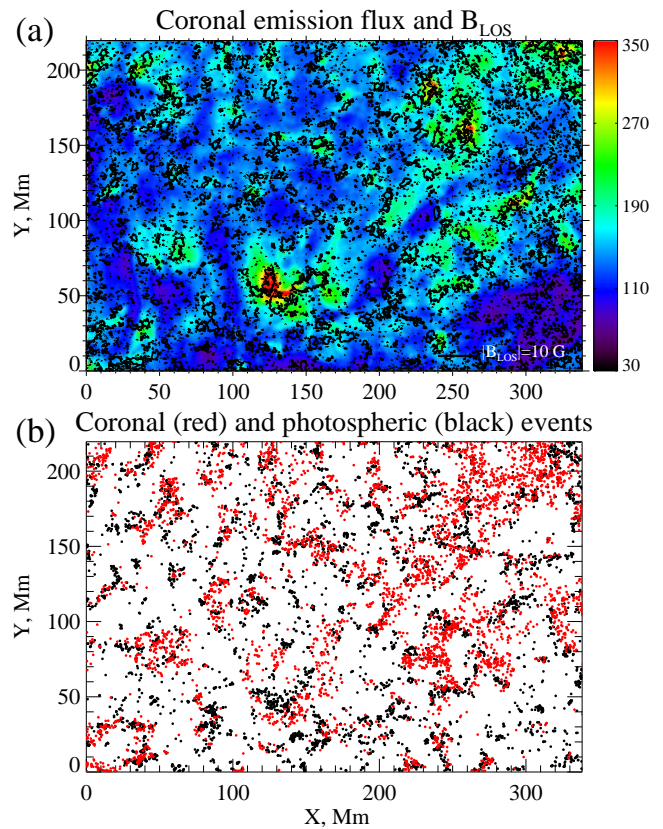}
\caption{\label{fig1} (a) Color-coded map of coronal emission flux (STEREO EUVI, 1:57:31 Apr 06 2007) superposed with contour lines of unsigned line-of-sight component of the photospheric magnetic field (SOHO MDI, 1:57:01 Apr 06 2007). (b) Starting positions of the coronal and photospheric events obtained using 95$\%$ and 99$\%$ percentile detection thresholds as explained in \cite{uritsky13}.}
\end{figure}

Fig. \ref{fig1}(a) shows a sample STEREO EUVI image combined with isolines (shown in black) of the unsigned line-of-sight SOHO MDI magnetic flux in the studied solar region. The observed fragmented magnetic carpet and sparse coronal emission pattern are typical of a quiet Sun \cite{leighton62, hagenaar97, shine00, rieutord10}. Some of the bright coronal locations coincide with magnetic field reversals, as expected for the quiet-Sun magnetic network (see, e.g., \cite{falconer98}), and could be sites of low-altitude magnetic reconnection \cite{karpen95}. 

The inter-event time of flaring events in the quiet solar corona is typically shorter than the event duration \cite{uritsky13}. Multiple events developing at different locations can overlap in time, and their proper analysis requires spatiotemporal \cite{berger96}, as opposed to time-series based \cite{boffetta99, paczuski05}, detection techniques. Our detection method \cite{uritsky02, uritsky07, uritsky10a} identifies image features staying for more than one sampling interval above a specified detection threshold and occupying separable connected subvolumes in the three-dimensional (3D) space-time. 

The first step of the applied feature-tracking technique consists of building a table of contiguous time intervals called activations where the local values of the studied data field exceeds the detection threshold \cite{uritsky10a}. Next, we labeled spatially connected clusters of activations using the ``breadth-first search'' principle to avoid backtracking of search trees. All 27 nearest neighbors in the 3D space-time, including the diagonal neighbors, were considered to identify the connected clusters which were treated as individual solar events. The detection thresholds were adjusted to represent comparable levels of intermittency in the studied image sets \cite{uritsky13}, yielding 4124 coronal and 5912 photospheric events (Fig. \ref{fig1}(b)). The results reported below for these sets of events have been also reproduced for several other combinations of thresholds. 
The occurrence time $t$ of every event was measured using two alternative methods: based on the event onset time $t_1$ or its average time $(t_1 + t_2)/2$, where $t_1$ ($t_2$) is the time of the first (last) image containing the event. We also recorded the average starting heliographic position $\mathbf{r}$ of each detected event. 

Due to a high average occurrence rate of coronal events (more than 4 events per sampling interval), their temporal clustering  could not be tested using inter-event time distributions \cite{boffetta99,lepreti01}. Instead, we applied the CI technique \cite{grassberger83} based on the analysis of second-order moments of the multifractal expansion of a clustered stochastic set \cite{grassberger83a, schuster05}. The CI characterizes the probability of finding a pair of data points within a hypersphere of a specified radius $\epsilon$ representing in our case temporal interval $\tau$ or spatial distance $r$, and is estimated by the sum
\begin{eqnarray}
C_{XY}(\epsilon) = K_{XY}  \sum_{\zeta_X \in X} \sum_{\zeta_Y \in Y}
\Theta \left( \epsilon - \left\|  \zeta_X - \zeta_Y \right\|  \right)
\label{eq1}
\end{eqnarray}
where $X, Y \in \{ H, M \}$ are sets of heating ($H$) or magnetic ($M$) events, $\zeta_{X,Y}$ are the occurrence times or the positions of the events in these sets, 
$K_{XY}$ is the normalization constant ensuring $C_{XY}(\epsilon) \to 1$ as $\epsilon \to \infty$, 
and $\Theta$ is the Heaviside step function. The combinations $\epsilon = \tau$,  $\zeta = t$ and $\epsilon = r$,  $\zeta = \mathbf{r}$ define respectively temporal and spatial CI. If the sets $X$ and $Y$ are identical, eq. (\ref{eq1}) describes auto-correlations between the events \cite{grassberger83}. For a self-similar clustered set $X$ of occurrence times and positions, $C_{XX}(\tau) \sim \tau^{\beta}$ and $C_{XX}(r) \sim  r^{\alpha}$, with $\beta \leq 1$ and $\alpha \leq 2$. The special cases $\beta = 1$, $\alpha = 2$ refer to the scaling of fully uncorrelated sets of events governed by the embedding dimension $d$ of the analysis domain. If the sets $X$ and $Y$ are distinct, the correlation dimensions characterize the closeness of two fractal measures \cite{kantz94}, with $\beta > 1$ ($\beta < 1$) and $\alpha > 1$ ($\alpha < 2$) signaling  negative (positive) cross-correlations in temporal and spatial domains, correspondingly \cite{uritsky12}. 

\begin{figure}
\includegraphics[width=8.5 cm]{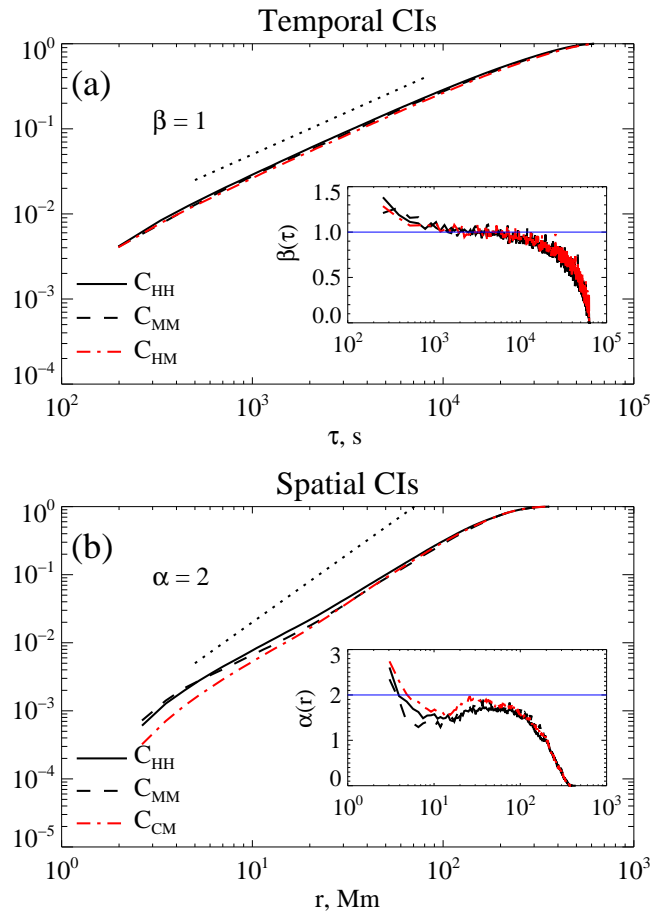}
\caption{\label{fig2} Analysis of temporal (a) and spatial (b) clustering of heating and magnetic events in the quiet Sun. Dotted lines represent log-log slopes expected in the absence of correlations. 
The insets show the dependence of local correlation dimensions on temporal and spatial scales revealing non-power law distortions eliminated in the next figure. }
\end{figure}

The obtained CI statistics of quiet Sun events are shown in Fig. \ref{fig2}. $C_{HH}$, $C_{MM}$, and $C_{HM}$ refer to the CIs characterizing respectively auto-correlations of heating events in the corona, magnetic events in the photosphere, and cross-correlations between the two. Temporal CIs are in an agreement with the non-clustering condition $\beta =1$ over an intermediate range of $\tau$ scales. As we demonstrate below, this range is considerably broader than the one seen in Fig. \ref{fig2}(a). Spatial CIs are described by $\alpha < 2$ at almost all $r$ scales, revealing clustered locations of coronal and photospheric events. Power-law slope of $C_{HM}(r)$ is also below 2 indicating positive cross-correlation between the positions of the events in these data sets. The physical coupling between the two solar regions is likely to be driven by free magnetic energy injected into the corona by vertical and horizontal photospheric convection \cite{welsch06}. The linear form of the temporal cross-CI $C_{HM} \sim \tau$ suggests that the occurrence times of heating events, unlike their positions, are statistically independent of the photospheric driver.

The local estimates of the correlation dimensions shown in the insets of Fig. \ref{fig2} diverge from their average values at the edges of the studied intervals of scales affected by the finite instrument resolution and the largest observable $\tau$ and $r$ values imposed by the size of the image and the duration of the image sequence. To eliminate these artifacts, 
we used the scaling ansatz $C(\epsilon) = \epsilon^{D} f(\epsilon) g(\epsilon) $, where $\epsilon$ is the scale of interest, $D$ is the corresponding correlation dimension, $f(\epsilon)$ is the cutoff function representing observational distortions, and $g(\epsilon)$ accounts for a non-power law scaling behavior intrinsic to the solar structure and dynamics. 
We performed a calibration based a synthetic set of events characterized by fully random timings and positions and subject to the same observational constraints as the solar events. 
The CI $\hat{C}(\epsilon$) of this random set of events is described by $g = 1$ and $D = d$, allowing us to introduce the corrected correlation dimensions
\begin{eqnarray}
D^*(\epsilon) = - \frac{\partial \,\mbox{log}\,[ C(\epsilon) / \hat{C}(\epsilon) ] }{\partial \,\mbox{log}\, \epsilon}  + d = D + \delta D(\epsilon),
\label{eq2}
\end{eqnarray}
in which $\delta D = \partial  \,\mbox{log}\, g(\epsilon) / \partial \,\mbox{log}\, \epsilon$ and $D^* \in \{\beta^*, \alpha^* \}$. By design, $D^*$ is independent of the observational distortions and approaches $D$ over a scaling range where $g$ is constant.

\begin{figure}
\includegraphics[width=8.5 cm]{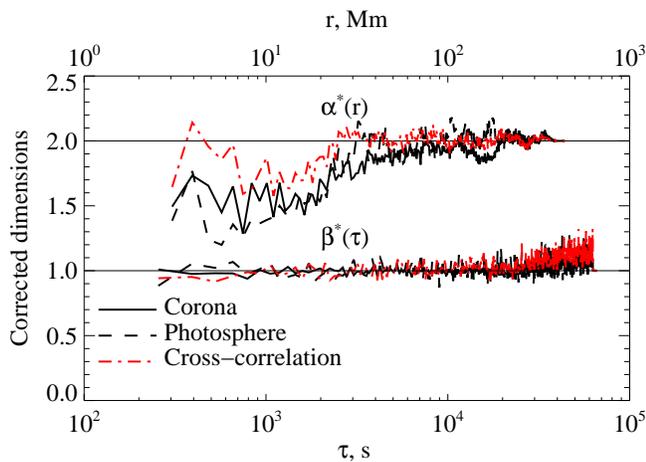}
\caption{\label{fig3} Corrected correlation dimensions $\alpha^*$ and $\beta^*$ revealing no significant temporal correlations up to $\tau \sim$ 15 hours but significant spatial clustering at the meso- and supergranulation scales. }
\end{figure}

Fig. \ref{fig3} displays the corrected dimensions $\beta^*$ and $\alpha^*$ estimated using eq. (\ref{eq2}) and plotted as a function of $\tau$ and $r$ scales. The average values and the standard errors of the corrected dimensions are provided in Table \ref{table1}. The data show that the temporal dimension is indistinguishable from the random prediction for $\tau = 3 \times 10^2 - 5 \times 10^4$ s. Within this range, the dynamics of the studied solar region is adequately described by the stationary Poisson process, in contrast to the active Sun showing fractal clustering of flaring times over an approximately the same range of scales \cite{aschwanden10}. 

The corrected spatial dimensions are consistently below the value 2 over the range of scales controlled by meso- and supergranulation (up to $r \sim 20 $ Mm) \cite{solanki06}. The leading role of the photosphere in the emergence of this spatial structure is manifested in its systematically lower $\alpha^*$ values (Fig. \ref{fig3}). The spatial cross-correlation dimension suggests that the positions of the coronal brightenings are arranged in a multsicale pattern which is consistent with the multiscale organization of the quiet Sun magnetic network \cite{rast03, uritsky12}. The multiscale co-alignment of the two event sets can be also seen in Fig \ref{fig1}(b).

\begin{table}
\caption{\label{table1} Corrected correlation dimensions}
\begin{tabular}{l | c  c | c c  }
\hline
 & $\beta^*$ 		& $\beta^*$ & $\alpha^*$ 	& $\alpha^*$ \\ 
 & onset times  	& aver. times & $r<20$ Mm 		& $r>20$ Mm		\\ 
\hline
$C_{HH}$   		&  1.01$\pm$0.03 &  1.00$\pm$0.02 & 1.56$\pm$0.12 & 1.97$\pm$0.08 \\ 
$C_{MM}$ 			&  1.03$\pm$0.07 &  1.00$\pm$0.03 & 1.45$\pm$0.14 & 2.01$\pm$0.06 \\ 
$C_{HM}$      &  1.06$\pm$0.07 &  1.02$\pm$0.03 & 1.78$\pm$0.15 & 2.01$\pm$0.04 \\ 
\hline
\end{tabular}
\end{table}



The absence of ensemble-averaged statistical correlations between the event timings leaves a possibility of a more subtle temporal organization associated with the presence of so-called sympathetic coronal eruptions \cite{torok11} triggering secondary instabilities via magnetohydrodynamic (MHD) waves. We tested this scenario using a causal network approach. Each coronal event was considered as a node of a directed graph. Outgoing links are added between a given event and all other events occurred within a 5 minute interval after that event, excluding the events described by a zero time lag. The links could represent causal connections between the events resulting in their nearly-simultaneous occurrence, but they can also be random. The Erd\H{o}s--R\'{e}nyi random graph \cite{erdos59} model predicts that in the latter case, the degree distribution $p(k)$ describing the probability of finding a node with $k$ incoming or outgoing links follows a binomial distribution which converges to a Poisson distribution for large $N$: $p(k) \approx e^{-\left\langle k \right\rangle} {\left\langle k \right\rangle}^k / k! $, where ${\left\langle k \right\rangle}$ is the average number of links per node \cite{erdos59, gilbert59, albert02}.

\begin{figure}
\includegraphics[width=8.5 cm]{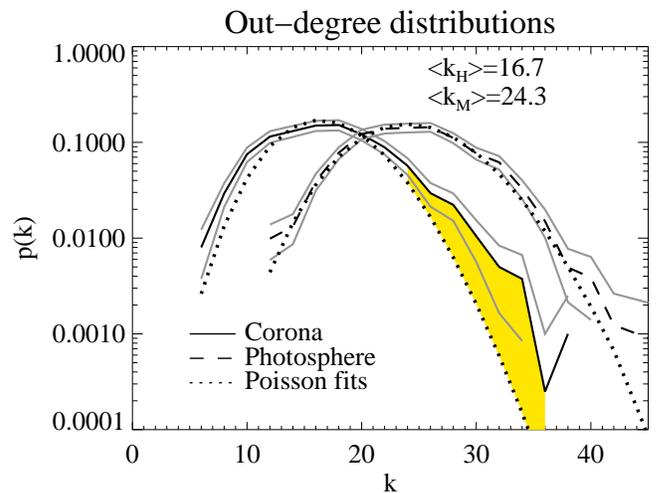}
\caption{\label{fig4} Out-degree distributions of the causal networks describing solar photosphere and corona.
Thin lines surrounding each distribution show histogram errors at the level of 3 standard deviations. Dotted lines are the Poisson fits corresponding to the Erd\H{o}s--R\'{e}nyi random graph model using the provided average numbers of links of heating ($\lambda_H$) and magnetic ($\lambda_M$) events.}  
\end{figure}

The out-degree distribution of the photospheric graph is fairly close to this prediction, but the coronal graph shows a systematic departure from the Poisson law for $k \gtrsim 25$ (Fig. \ref{fig4}) where the number of outgoing links is systematically larger compared to the random graph. These hub events have no effect on the CI shape as they account for less then 3$\%$ of the total population of events, but they can be quite important as triggers of secondary heating activity. Sympathetic events in solar active regions have been considered in the context expanding flux ropes working as MHD triggers of a second generation of reconnection events \cite{torok11}. Our analysis speaks in favor of such remote interactions in the quiet corona. We found that the average transverse coronal length of the outgoing links attached to highly-connected events is $160 \pm 30$ Mm. This distance is traveled in $\leq 300$ s, implying a communication speed of 500 km/s or faster. The provided estimate is consistent with the Alfv\'{e}n speeds in the corona measured using the coronal magnetography \cite{brosius97} and other remote techniques (see e.g. \cite{warmuth05} and refs therein). 

Together with the power-law distribution of events sizes published earlier \cite{uritsky13}, our findings indicate that the bulk heating of the corona can be controlled by SOC-like avalanches of energy dissipation \cite{bak87, lu91} up to the time scales comparable with the lifetime of supergranulation cells \cite{rieutord10}. The preferred locations of heating events are preconditioned by the nonuniform distribution of free magnetic energy supplied by the photosphere \cite{falconer98}. However, the moments at which the stored energy is released are likely to be determined by local instability conditions rather than the large-scale energy supply, and are essentially random. A more coherent energy dissipation associated with turbulent plasma heating (see e.g. \cite{buchlin07}) could, in principle, occur at $\sim 10^2$ s time scales not properly resolved in this study. Finally, the performed causal network analysis suggests that bulk coronal heating is, at least partially, a nonlocal phenomenon which may involve long-distance Alfv\'{e}nic interactions between remote coronal regions. This non-locality could play a significant part in triggering secondary instabilities in the quiet corona and must be addressed in future models of coronal heating.

\begin{acknowledgments}
We acknowledge J. Klimchuk for helpful comments, A. Coyner for preparing solar images. The work of V.U. was supported by NASA Grant NNG11PL10A 670.002 to CUA / IACS.
\end{acknowledgments}

\bibliography{apj_2010}

\end{document}